\documentclass[conference]{IEEEtran}
\usepackage{cite}
\usepackage[bottom=2cm, right=2cm, left=2cm, top=1.2cm]{geometry}
\usepackage{amsmath,amssymb,amsfonts}
\usepackage{amsthm}
\usepackage{algorithmic}
\usepackage{graphicx}
\usepackage{textcomp}
\usepackage{physics}
\usepackage{xcolor}
\usepackage{mathtools}
\usepackage{tabularx}

\def\BibTeX{{\rm B\kern-.05em{\sc i\kern-.025em b}\kern-.08em
    T\kern-.1667em\lower.7ex\hbox{E}\kern-.125emX}}

    \newtheorem{definition}{Definition}

    \newtheorem{theorem}{Theorem}
    
    \newtheorem{example}[theorem]{Example}
    
\newcounter{protocol}
\newenvironment{protocol}[1]
  {\par\addvspace{\topsep}
   \noindent
   \tabularx{\linewidth}{@{} X @{}}
    \hline
    \refstepcounter{protocol}\textbf{Protocol \theprotocol} #1 \\
    \hline}
  { \\
    \hline
   \endtabularx
   \par\addvspace{\topsep}}

\begin{document}

\title{Unconditional Proofs-of-Work and Other Possibilities of Thermodynamic Cryptography\vspace{-3mm}}

\author{\IEEEauthorblockN{Xavier Coiteux-Roy$^{1,2}$ and Stefan Wolf$^1$}
	\IEEEauthorblockA{$^1$\textit{Universit\`a della Svizzera italiana},
Lugano, Switzerland.\\$^2$\textit{Technische Universit\"at M\"unchen}, Germany. \\
\{xavier.coiteux.roy, stefan.wolf\}@usi.ch}
}

\maketitle

\begin{abstract}
In line with advances in recent years about realizing cryptographic functionalities in an information-theoretically secure way from physical phenomena and laws, 
we propose here to obtain useful tasks from the sole assumption of limited free energy. 
Specifically, based on that assumption~--- resulting in a setting loosely related to Maurer's bounded-storage model~---
we derive protocols for unconditional proofs-of-thermodynamical-work, secret sharing of free energy, unforgeable money, and proofs-of-position. While our schemes can be considered classical and not quantum per se, they are resistant against both classes of adversaries.
\end{abstract}

\section{Introduction}

\subsection{Cryptographic assumptions}

In all the different paradigms and frameworks considered in the history of cryptography, 
it has been an unshakeable constant that cryptographic security --- be it secrecy, 
authenticity, or the possibility of co\"operation without trust --- is
based on some ultimate assumption. This can be the availability of shared secret pieces of
information, called keys, the hardness of certain computational problems with respect 
to classical or quantum Turing machines (or any other model of computation), 
or a physical phenomenon or law. 

Generally, the underlying assumption translates to, or can be rephrased as, some 
sort of limitation on the adversary's power and capabilities, e.g., in terms of her accessible
information, computational resources, or memory space. Of particular interest is 
{\em information-theoretic} --- as opposed to merely computational --- security   
derived from physical effects and laws, such as noise in communication channels~\cite{wyner1975wire,csiszar1978broadcast,ahlswede1993common,maurer1993secret}, the uncertainty principle of quantum physics~\cite{bb84,ekert1991quantum}, Bell nonlocal correlations combined with the no-signalling principle of special 
relativity~\cite{barrett2005no,hanggi2010efficient,masanes2014full}; on top of that, the latter {\em alone\/}
allows for bit commitment~\cite{kent1999unconditionally}~--- which is {\em im\/}possible to achieve from quantum
physics only~\cite{mayers1997unconditionally}. 

We propose in the present work to start from the sole assumption that the opponent's accessible 
{\em free energy\/} is limited. We present protocols for cryptographic functionalities such as 
proofs-of-work, secret sharing of free energy, unforgeable money, and secure positioning.

\subsection{The second law of thermodynamics}
The second law of thermodynamics, in one of its formulations, states that in a closed system, 
{\em entropy does not decrease} (Boltzmann). This implies, in particular, that {\em free energy\/} cannot be 
created out of thin air, i.e., from {\em one single\/} heat bath (Kelvin). Equivalent to such traditional readings 
of the law is {\em Landauer's principle\/}: The erasure of information has a thermodynamic price~\cite{landauer61,bennett1982thermodynamics,bennett2003notes,faist2015minimal}.
More specifically, it states that erasing~$N$ bits of information\footnote{More precisely in the quantum case, $N$ bits of max-entropy.} requires free energy of\footnote{Here, $k_B$ is Boltzmann's constant and $T$ is the absolute temperature of the environment.}
$k_BT\ln 2\cdot N$, which is then dissipated as heat into the environment in order to compensate
for the entropy defect in the memory device linked to the information erasure. Assuming that a 
possible adversary's access to free energy be limited then translates via Landauer's principle 
to a limit on the opponent's ability to access a memory initiated by the all-$0$-string (or any pure quantum state that acts as a reference); the overall 
memory capacity in turn need not be bounded in our model, which makes both players and adversaries stronger than in the 
well-studied "bounded-storage model"~\cite{maurer1992conditionally,cachin1998oblivious,dziembowski2002tight,ding2004constant,dziembowski2004generating,damgaard2008cryptography,konig2012unconditional}. We propose here unconditional, i.e., 
information-theoretically secure, versions of prominent cryptographic functionalities such as 
unforgeable money and secure positioning, and we introduce natural tasks that we call \emph{proofs-of-thermodynamical-work} and \emph{secret sharing of free energy}.

\subsection{The model}
Each player (Alice, Bob, Prover, Verifier, etc.) is modeled as a powerful non-uniform quantum (or classical\footnote{Note that all of our protocols are \emph{classical} in the sense that the honest players never need to make truly quantum operations (they can act as if all involved density matrices were diagonal). We use the quantum-information notation because our protocols are, however, resistant to quantum adversaries that can apply any unitary quantum operation. The security relies only on the fact common to both classical and quantum mechanics that their laws of motion are reversible --- i.e., information is never lost.}) circuit whose only restriction is its free-energy consumption. That is, it has unbounded computational power in the complexity sense and can act on arbitrarily large memories as long as it acts in a logically reversible manner (i.e., it can do any unitary operation in a thermodynamically reversible manner, so for free). One task that it cannot, however, do for free is, mainly, to erase arbitrarily large amounts of information (where erasure means a reset to the all-$\ket{0}$ ground state).

We consider two kinds of registers for these players~--- their sizes are defined asymptotically in some parameter $n$.
\begin{enumerate}
	\item A bounded memory to emulate the free-energy consumption of the player. This register initially contains a blank state (i.e.,~${\ket{0}^{\otimes f(n)}}$ , the length $f(n)$ of which depends on the class of the player (see Section~\ref{section-classes}). It is used implicitly to keep track of the thermodynamical work made by each of the players --- they are restricted to reversible quantum operations (unitary) and might draw ancillas from that memory to simulate more general irreversible quantum operations (CPTP maps). This blank memory is also used to encode the linear amount of classical information necessary to the protocols.
	\item Large unbounded memories represented by (sometimes partial) density matrices written as capital letters $X$ or $Y$. The registers can be exchanged between the players, as explained below. They are considered to float in the environment and they have the specificity of always being initialized as uniformly random mixed states (thermodynamical equilibrium) $\pi_N:=\sum_{i\in[0,2^N-1]}\frac{\ketbra{i}}{2^N}$, where $N$ will be specified, but can be a function of $n$ such as the exponential or ${\operatorname{Ackermann}(n)}$. The sizes of the associated complex Hilbert space $\mathcal{X}$ and $\mathcal{Y}$ are taken as implicit.
\end{enumerate}

We supplement those memories with two modes of communication.  
\begin{enumerate}
	\item Standard communication, which can be copied and broadcast, but that is limited to a linear amount $O(n)$. We use the term ``to tell'' for this type of communication.
	\item Physical channels for sending the unbounded memories from one player to another. The process is considered to be thermodynamically neutral and is not limited by the size of the memories, but the sender retains no copy of what they send. We use the term ``to send'' for this type of communication. Formally, this channel is a swap: $\mathcal{X}_A \otimes \mathbb{C}^0_B \rightarrow \mathbb{C}^0_A \otimes \mathcal{X}_B$.
\end{enumerate}
\subsection{Free-energy bounded players}\label{section-classes}
In this work, we analyze various classes of players according to their asymptotic bound in free energy (which corresponds to the length in qubits of their initially blank memory~---~we set $k_B T \ln 2 :=1$).
We concern ourselves with three classes: $f(n)$ is linear; $f(n)$ is exponential; and $f(n)$ is an arbitrary computable function that is growing extremely fast, such as $\textrm{Ackermann}(n)$. We specify a player's bound in subscript and write for example Verifier$_{\textrm{lin}(n)}$ or Prover$_{f(n)}$.

\subsection{Proofs-of-thermodynamical-work}
We introduce the concept of proof-of-work, where \emph{work} is to be taken in its literal, thermodynamic sense: as expenditure of free energy.

Proofs-of-thermodynamical-work discriminate the different classes of players. We will mainly be concerned with statements asymptotic in the parameter $n$.
\begin{definition}\rm
A proof-of-thermodynamical-work of value $f(n)$ is a state~$X=\ketbra{0}^{f(n)}$, where $f(n)$ is a computable function.
\end{definition}

An information-theoretic proof-of-thermodynamical-work can be used in the following, straightforward protocol to allow Prover$_{f(n)}$ to convince Verifier$_{\textrm{lin}(n)}$ that they really possess (except with probability exponentially small in $s$) an amount $f(n)-s$ of free energy, where $s$ is a security parameter. 
The function $f(n)$ is arbitrary, but we require that $f(n)\in \omega(n)$. It could be for example quadratic, exponential, or the Ackermann function. 
\begin{protocol}{Unconditional proof-of-work of amount $f(n)$}
\textbf{Prover}$_{\mathbf{f(n)}}$\hfill\textbf{Verifier}$_{\mathbf{lin}(n)}$\\
$X\leftarrow\pi_{f(n)}:={{\sum_{i\in[0,2^{f(n)}-1]}\frac{\ketbra{i}}{2^{f(n)}}}}$\hfill\\
$X\xrightarrow{\text{work is done to erase}} \ket{0}^{f(n)}$ \hfill\\

$\ket{0}^{f(n)}\xrightarrow{\hspace{2cm}\text{Proof-of-work is sent}\hspace{3cm}}$\hfill~
\end{protocol}

The perfect completeness of the above protocol is straightforward: If the prover possesses an amount of at least $f(n)$ of free energy, it can use it for work, that is, to reset a state of $f(n)$ qubits to the all-$\ket{0}$ string.
\begin{theorem}[Information-theoretic soundness]\label{theoremwork}
No Prover$_{g(n)}$ with $g(n)\le f(n)-s$ can produce a proof-of-thermodynamical-work of value $f(n)$ except with exponentially small probability $2^{-s}$.

\begin{proof}
In \cite{crypto2022}, a full proof of the statement regarding classical adversaries is given using Turing-machine formalism (uniform circuits). We sketch here a proof valid in presence of quantum adversaries. Let an adversary execute on its starting resources the arbitrary unitary operation $U(\ketbra{0}^{g(n)}\otimes \pi_t)U^\dagger$ with $\pi_t=I_t/2^t$ a maximally mixed state of $t$ qubits (with $t$ arbitrary).  It holds that
\begin{align*}
&\tr \left( (\ketbra{0}^{g(n)+s}\otimes I_{t-s})\cdot U(\ketbra{0}^{g(n)}\otimes \pi_t)U^\dagger \right)\\
&=2^{-s}\tr \left(U (\ketbra{0}^{g(n)+s}\otimes \pi_{t-s})U^\dagger\cdot (\ketbra{0}^{g(n)}\otimes I_t) \right)\\
&\le 2^{-s}\,.
\end{align*}

The physical interpretation is that creating $\ket{0}$s out of thin air (or out of completely mixed states) with higher than $2^{-s}$ probability would allow via Szil\'ard's engine~\cite{szilard1929entropieverminderung,rio2011thermodynamic} for a perpetuum mobile of the second kind and violate the second law of thermodynamics.
\end{proof}
\end{theorem}
In this work, we construct protocols for different tasks in this described framework of ``thermodynamical cryptography''.

\section{Non-transferable interactive proofs-of-thermodynamical-work}

In the straightforward protocol given previously, the prover convinces the verifier of their work by giving them directly the blank state $\ket{0}^{f(n)}$. This enables, of course, the verifier to convince a third party --- for example, a judge --- that they themselves (the verifier) have produced the work. This motivates us to look at \emph{non-transferable} proofs-of-thermodynamical-work, achievable through interaction, which in spirit resemble zero-knowledge proofs. 

\begin{definition}[Non-transferable proof-of-work]
A protocol $(P,V)$ achieving a proof-of-thermodynamical-work is perfectly non-transferrable if for any Verifier $\hat{V}$, there exists a Simulator $S_{\hat{V}}$, with the same bound in free energy as the verifier, that can reproduce the state of the memory of $\hat{V}$ at the end of the interaction $(P \leftrightarrow \hat{V})$. In other words, the bound in free energy of any (possibly malicious) Verifier is not allowed to increase through its interaction with the Prover.
\end{definition}
We present two protocols that achieve this task with perfect completeness (powerful enough provers always convince the verifier) and information-theoretic soundness (underpowered provers fail except with exponentially small probability). We use the protocols later as building blocks to achieve other cryptographic tasks.

\subsection{Non-transferable proof-of-work (sampling)}

The first protocol uses a technique based on sampling states of size exponential in $n$. The argument was introduced in~\cite{crypto2022} in the context of secret-key establishment and oblivious transfer. We limit ourselves to strings of exponential size to make broadcasting the sampling positions possible, but, as we see in Section~\ref{sectionhashing}, the restriction can be circumvented to accommodate more powerful provers. In the following, a honest prover turns two independently mixed states $(\pi_{2^{n+1}} \otimes \pi_{2^{n+1}})$ into two identical copies $\hat{X}_{AB}=\sum_{i\in[0,2^{n+1}-1]}\ketbra{i,i}$ of a mixed state.

\begin{protocol}{Non-transferable proof-of-thermodynamical-work (sampling)}
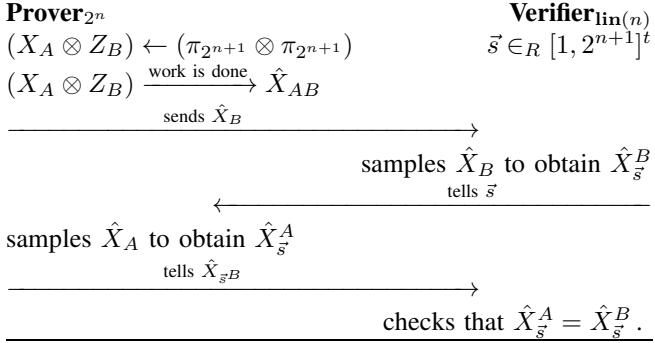
\textbf{Prover}$_{{2^{n}}}$ \hfill \textbf{Verifier}$_{\mathbf{lin}(n)}$\\
$(X_A \otimes Z_B) \leftarrow (\pi_{2^{n+1}} \otimes \pi_{2^{n+1}})$ \hfill $\vec{s}\in_R[1,2^{n+1}]^{t}$\\
$(X_A \otimes Z_B)\xrightarrow{\text{work is done}} \hat{X}_{AB}$  \\
$\xrightarrow{\hspace{2cm}\text{sends~}\hat{X}_B\hspace{3cm}}$\\
\hfill samples $\hat{X}_B$ to obtain ${\hat{X}}_{\vec{s}}^B$\\
\hfill$\xleftarrow{\hspace{3cm}\text{tells~}\vec{s}\hspace{2cm}}$\\
{samples $\hat{X}_A$ to obtain} ${\hat{X}}_{\vec{s}}^A$\hfill\\
$\xrightarrow{\hspace{2cm}\text{tells~}\hat{X}_{\vec{s}^B}\hspace{3cm}}$\\
\hfill {checks that} ${\hat{X}}_{\vec{s}}^A=\hat{X}_{\vec{s}}^B$\,.
\label{protocolsampling}
\end{protocol}
The protocol is perfectly complete (note that Verifier$_{\textrm{lin}(n)}$ cannot sample a state of more than exponential length). The proof-of-work is also non-transferable for reasons similar to Theorem~\ref{nt}. We prove the soundness.
\begin{theorem}[Information-theoretic soundness]\label{theoremwork2}
In Protocol~2, no Prover$_{2^n}$ can asymptotically produce a proof-of-thermodynamical-work of value $2^{n+1}$ except with probability exponentially small in $\min(n,t)$, where $t$ is the number of samples.
\begin{proof}
We sketch the proof (see~\cite{crypto2022} for details). Using Proposition~3 of~\cite{crypto2022}, we find that $H_\infty(\hat{X}^B|E)\ge 2^n$, where $E$ is all of the side information of Prover$_{2^n}$. Since sampling approximately preserves the min-entropy per bit in presence of classical side information~\cite{vadhan2004constructing}, as well as in presence of quantum side information~\cite{konig2011sampling}, it holds that $H_\infty(\hat{X}_{\vec{s}}^B|E)\gtrsim 2^n \cdot t/2^{n+1}=t/2$ with overwhelming probability in the size $t$ of the sample $\vec{s}$. We conclude that $P({X}_{\vec{s}}=\hat{X}_{\vec{s}}) \lesssim 2^{-t/2}  $ for any Prover$_{2^n}$.
\end{proof}
\end{theorem}

\subsection{Non-transferable proof-of-work (hashing)}\label{sectionhashing}
We introduce a variant based on a novel method using universal hashing that allows Verifier$_{\textrm{lin}(n)}$ to authenticate, despite their linear free-energy limitation, arbitrarily long strings. This method has two merits: First, the results using universal hashing are tighter than the ones using sampling; Second, using an arbitrarily long string $Y$ as a seed allows for non-transferable proofs-of-thermodynamical-work whose amounts are any function~$f(n)$, even, e.g., the Ackermann function. (Again, $\hat{X}_{AB}$ represents here two perfectly correlated copies of a completely mixed state.)

\begin{protocol}{Non-transferable proof-of-thermodynamical-work (hashing)}
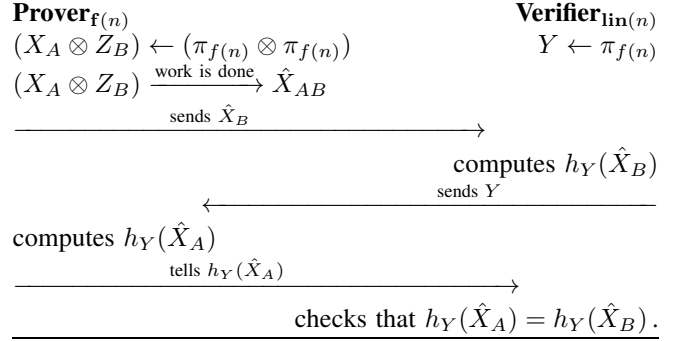
\textbf{Prover}$_{\mathbf{f}(n)}$ \hfill \textbf{Verifier}$_{\mathbf{lin}(n)}$\\
$(X_A\otimes Z_B)\leftarrow ( \pi_{f(n)}\otimes \pi_{f(n)})$ \hfill $Y\leftarrow \pi_{f(n)}$\\
$(X_A\otimes Z_B)\xrightarrow{\text{work is done}} \hat{X}_{AB} $  \\
$\xrightarrow{\hspace{2cm}\text{sends~}\hat{X}_B\hspace{3cm}}$\\
\hfill computes ${h}_Y(\hat{X}_B)$\\
\hfill$\xleftarrow{\hspace{3cm}\text{sends~}Y\hspace{2cm}}$\\
{computes} ${h}_Y(\hat{X}_A)$\hfill\\
$\xrightarrow{\hspace{2cm}\text{tells~}h_Y(\hat{X}_A)\hspace{3cm}}$\\
\hfill {checks that} ${h}_Y(\hat{X}_A)={h}_Y(\hat{X}_B)$\,.
\end{protocol}
Here, $h_Y(\cdot): \{0,1\}^{f(n)} \times  \{0,1\}^{f(n)} \rightarrow \{0,1\}^t$ is a family of universal hashing functions~\cite{carter1979universal,wegman1981new}, whereas $Y$ is the seed for the choice of the hash function. Note that this process necessitates only a quantity $t$ of free energy (of $\ket{0}$s).
\begin{theorem}[Information-theoretic soundness]\label{hashingproof}
In Protocol~3, no Prover$_{g(n)}$ with $g(n)\le f(n)-s$ can produce a proof-of-work of value $f(n)$ except with probability exponentially small in $\min{(s,t)}$.
\begin{proof}
We first remark that $h_Y(\hat{X}_A)=h_Y(\hat{X}_B)$ implies that $\hat{X}_A=\hat{X}_B$ except with probability $2^{-t}$, because of the universality of the hashing~\cite{bennett1988privacy,impagliazzo1989pseudo,haastad1993construction,bennett1995generalized}. Since the prover can do the operation $\hat{X}_{AB}\rightarrow \pi_A \otimes \ketbra{0}^{f(n)}$ logically and thermodynamically reversibly, the problem reduces to the one of Theorem~\ref{theoremwork}.
\end{proof}
\end{theorem}

\begin{theorem}[Perfect non-transferability]\label{nt}
The proof-of-thermodynamical-work achieved by Protocol~3 is perfectly non-transferable.
\begin{proof}
Since the simulator does not need to copy the system~$X$ to simulate its access by both the verifier and the prover, it can simulate the protocol perfectly.
\end{proof}
\end{theorem}

\section{Secret sharing of free energy}
We remark that at the end of Protocol~3, the prover and the verifier, whom we now rename Alice and Bob, share a correlated state $\hat{X}_{AB}=\sum_{i\in[0,2^{f(n)}-1]}\ketbra{i,i}$. Locally, this state is completely mixed and does not enable Alice or Bob to produce work --- it has no free-energy value. However, if Alice and Bob were to meet and join their states together, they could do the compressing operation $\hat{X}_{AB}\rightarrow \pi_A \otimes \ketbra{0}^{f(n)}$ to produce the state $\ket{0}^{f(n)}$ and, therefore, extract a corresponding extra amount $f(n)$ of free energy. In some sense, this is the free-energy analogue of secret sharing. The concept generalizes to any bipartite state.
\begin{theorem}[Secret sharing of free energy (bipartite)]
Any joint quantum state $X_{AB}$, belonging to Alice and Bob, respectively, constitutes a secret sharing of free energy of amount $H_0(X_A)+H_0(X_B)-H_0(X_{AB})$.
\begin{proof}
This difference in min-entropy is the additional compression in the worst case that appears when the two systems are reunited. The freed space is used as a proof-of-work.
\end{proof}
\end{theorem}
\noindent This concept can be extended to multipartite states.
\begin{example}[Secret sharing of free energy (multipartite)]
Let $X_{A_1,\dots,A_k}$ be a quantum state of dimensions $2^{kN}$, split among Player $1$ to $k$, respectively, with entropy maximal given the constraint $X_{A_1}\oplus\dots \oplus X_{A_k}=\ketbra{0}^N$. This redundant state constitutes a $k$-partite secret sharing of free energy $N$.\end{example}

\section{Unforgeable money}
\subsection{Unforgeable banknotes (sampling)}
The concept of unforgeable money was introduced by Wiesner in \cite{wiesner1983conjugate} in the context of quantum information. A~similar construction can be made through free-energy bounds. We propose unforgeable banknotes that can be used $k-1$ times before a trip to the bank be necessary.
\begin{protocol}{Unforgeable banknotes (sampling)}
\noindent \textbf{Emission}\\
\noindent To emit banknote \#$i$, the bank samples $X_i = \pi_{2^{n+1}}$ at $k\cdot t$ positions and then physically transfers the state $X_i$ into the banknote.\\
\noindent \textbf{Spending} (up to $k-1$ times)\\
\noindent To verify the legitimacy of banknote \#$i$, the merchant calls the bank and asks them for sampled positions and values; if it is the first inquiry about bill $\#i$, the bank gives the first sample (of size $t$); if it is the second inquiry, the bank gives the second sample, and so on, until the $(k-1)^\textrm{th}$ time, when it warns that a trip to the bank be necessary to re-emit the banknote.\\
\noindent \textbf{Re-emission}\\
The $k^\textrm{th}$ sample is used by the bank that then re-emits the bill. If the samples do not match the bill, the bill is invalid and refused by the merchant (or bank).
\end{protocol}

\begin{theorem}
The probability that such a banknote be successfully forged by an Adversary$_{2^n}$ is exponentially small in $\min(n,t)$.
	\begin{proof}
Note that every round of sampling is independent and that the statement thus reduces to Theorem~\ref{theoremwork2}.\end{proof}
\end{theorem}

\subsection{Unforgeable tickets (hashing)}
A variant of the unforgeable banknotes are the unforgeable tickets, where the verifier is also the emitter. We give a protocol that is secure against opponents with arbitrarily fast asymptotically growing free energy.
\begin{protocol}{Unforgeable concert tickets (hashing)}
\noindent \textbf{Emission}\\
\noindent To emit ticket \#$i$, the concert venue takes $(X_i \otimes Y_i)= \pi_{f(n)}\otimes \pi_{f(n)}$ and computes the value of the universal hash function $h_Y(X): \{0,1\}^{f(n)} \times  \{0,1\}^{f(n)} \rightarrow \{0,1\}^t$. It then physically transfers the state $X_i$ into the ticket, and stores $Y_i$ and $h_{Y_i}(X_i)$.\\
\noindent \textbf{Entrance}\\
\noindent To verify the legitimacy of ticket \#$i$, the merchant extracts~$\hat{X_i}$ from the ticket and computes the hash $h_{Y_i}(\hat{X}_i)$. If it matches the recorded $h_{Y_i}(X_i)$, the merchant recognizes the ticket; otherwise, they reject it.
\end{protocol}

\begin{theorem}
The probability that such a ticket be successfully forged by an Adversary$_{g(n)}$ with $g(n)\le f(n)-s$ is exponentially small in $\min{(s,t)}$.	\begin{proof}
The proof is similar to the one of Theorem~\ref{hashingproof}.
	\end{proof}
\end{theorem}

\section{Position-based cryptography}
\subsection{Secure positioning (sampling)}
As a further application of our model, we turn to position-based cryptography~\cite{chandran2009position,buhrman2014position}. More precisely, we present protocols for secure positioning, that is, schemes for two\footnote{As a proof of principle, we use two verifiers. Two verifiers are enough if the timing of the prover is perfect (i.e., instantaneous computation and speed-of-light communication) and if he is exactly in the middle, but we note that in general, a third verifier is necessary to establish a position through triangulation.} verifiers to assert that the prover is at a specific geographical point. These proofs-of-position assume the validity of special relativity, that is, of a limit on the speed of light. We start with a first protocol based on sampling. It does not require the large-entropy physical system to travel at a speed close to the speed of light; only the standard communication must.

\begin{protocol}{Secure positioning (sampling)}
The two verifiers share a set of $t$ positions $\vec{s}\in[1,2^{n+1}]^t$ and a one-time pad $\vec{a}\in_R [1,2^{n+1}]^t$ (both are classical information). The verifiers are at a distance~$d$ and verify that the prover is exactly at the unique point~$P$, which is at a distance~$d/2$ from both.\\

$\triangleright$ Verifier~1 starts by generating $X=\pi_{2^{n+1}}$ from their environment and samples it according to $\vec{s}$, obtaining $X_{\vec{s}}$. It shares $X_{\vec{s}}$ with Verifier~2.\\
~

$\triangleright$ \hspace{-0.25cm}\hfill\textbf{Verifier 1} \hfill \textbf{Prover} \hfill \textbf{Verifier 2}\\
$t<0: \hfill\xrightarrow{\text{sends~}X\hspace{1cm}}\hfill\hfill~$\\
$t=0:\hfill\hfill \xrightarrow{\textrm{tells~} \vec{a}\hspace{1.5cm}} \hfill \xleftarrow{\textrm{tells~} \vec{b}=\vec{a}+\vec{s}\mod 2^n}\hspace{0cm} \hfill~$\\
$t=d/c :\hfill \xleftarrow{\hspace{1cm}\textrm{tells~} \hat{X}_{\vec{s}}}~~ \xrightarrow{\textrm{tells~} \tilde{X}_{{\vec{s}}}\hspace{1cm}}\hfill~$\\
~

$\triangleright$ Finally, the verifiers check that $X_{\vec{s}}=\hat{X}_{\vec{s}}=\tilde{X}_{\vec{s}}$, and that they both received the answer in time $d/(2c)$. If it is the case, they accept the proof-of-position; otherwise, they reject it.
\end{protocol}

\begin{theorem}
The above protocol for secure positioning is information-theoretically secure against Prover$_{2^n}$, i.e., such a prover cannot convince, with more than probability exponentially small in $t$, the verifiers that someone is at point~$P$ if neither she nor its potential agents are there.
\begin{proof}
The soundness of Protocol~2 guarantees that if $X_{\vec{s}}=\hat{X}_{\vec{s}}=\tilde{X}_{\vec{s}}$, then the two sampling procedures must have been done on the same system (not on two copies), unless with probability exponentially small in $t$. Since there is exactly one position in space where both sampling procedures could have been done in time to answer to the respective verifiers, the prover must be exactly there, at the centre of the line going from Verifier~1 to Verifier~2.
\end{proof}
\end{theorem}

\subsection{Secure positioning (hashing)}
We continue with a variant using hashing that is robust against more powerful provers (such as Prover$_{\textrm{Ackermann}(n)}$). It requires, however, that the prover be able to send states of dimension $f(n)$ at the speed of light.
\begin{protocol}{Secure positioning (hashing)}
The protocol uses $r$ rounds composed in parallel. The following describes one round.
The verifiers are at a distance~$d$ and verify that the prover is exactly at the unique point~$P$, which is at a distance~$d/2$ from both.\\
~

$\triangleright$ Verifier~1 starts by generating $X\otimes Y=\pi_{f(n)}\otimes\pi_{f(n)}$ from their environment and computes the value of the universal hash function $h_Y(\cdot): \{0,1\}^{f(n)} \times  \{0,1\}^{f(n)} \rightarrow \{0,1\}^t$. The verifiers share random bits $a\in_R\{0,1\}$ and $b\in_R\{0,1\}$. Verifier~1 sends the state $Y$ to Verifier~2 if $a\oplus b=1$.\\
~

$\triangleright$ \hspace{-0.25cm}\hfill\textbf{Verifier 1} \hfill \textbf{Prover} \hfill \textbf{Verifier 2}\\
$t<0: \hfill\xrightarrow{\text{sends~}X\hspace{1cm}}\hfill\hfill~$\\
$t=0:\hfill\hfill \xrightarrow{\textrm{tells~} {a}\hspace{1.5cm}} \hfill \xleftarrow{\hspace{1.5cm}\textrm{tells~} {b}}\hfill~$\\
$t=d/c :\hfill \xleftarrow{\textrm{sends~} \hat{X}\textrm{~if~} a\oplus b =0 }~~ \xrightarrow{\textrm{sends~} \tilde{X}\textrm{~if~}a\oplus b =1}\hfill~$\\
~

$\triangleright$ Finally, the two verifiers share their results. If $a\oplus b=0$ and Verifier~1 measures $h_Y(X)=h_Y(\hat{X})$, or if $a\oplus b =1$ and Verifier~2 measures $h_Y(\tilde{X})$; and if they both received the answer at time $d/c$, the verifiers accept the round. If all $r$ rounds are accepted, the proof-of-position is recognized; otherwise, it is rejected.\end{protocol}

\begin{theorem}
The above protocol for secure positioning is information-theoretically secure against Prover$_{g(n)}$ with $g(n)\le f(n)-s$, i.e., such a prover cannot convince the verifiers, with probability non-negligible in $\min{(r,s,t)}$, that someone is at point~$P$ if neither her nor her potential agents are there.
\begin{proof}
We start with the argument of Theorem~\ref{hashingproof} based on universal hashing: Here, $h_Y(X)=h_Y(\hat{X})$ implies that $X=\hat{X}$ except with probability $2^{-t}$ if $a\oplus b =0$, and \emph{idem} for $h_Y(X)=h_Y(\tilde{X})$, implying that $X=\tilde{X}$ if $a\oplus b =1$.

We then remark that if after $r$ rounds, the prover has always been able to send $X$ in time $d/(2c)$ to the prover designed by the value of $a\oplus b$, it must be that they can do so independently of the secret value of $a\oplus b$ determined randomly and secretly at the beginning of the protocol.

But we remark that a Prover$_{g(n)}$ with $g(n)\le f(n)-s$ cannot copy $X$ such that $X=\hat{X}$ and $\tilde{X}=X$ be encoded in two distinct but perfectly correlated systems. They must originate from the same point in space, and thus right from the centre of the line going from Verifier~1 to Verifier~2.

Finally, if the prover cannot send $X$ to both provers independently of the value of $a\oplus b$, they have probability at least~$1/2$ of being caught each round.
\end{proof}
\end{theorem}

\section{Conclusion}
We propose the study of cryptographic functionalities based on facts from thermodynamics. This research is in line with, and continues, the paradigm of basing information-theoretic security on physical laws.

While simple, our model of ``thermodynamic cryptography'' is rich in its cryptographic possibilities. In an environment where matter and entropy (i.e., the quantity of information) come in very large amounts, but where free energy is severely limited, an \emph{almost-no-cloning principle} emerges, different in its mechanics from the quantum no-cloning principle: Most of the information of large systems must reside at a unique point in space-time, while the monogamy of entanglement, for example, allows the ``information'' of the outcome of identical measurements of a maximally entangled quantum state to be obtainable at two points in space.

We propose schemes for realizing cryptographic functionalities, such as proofs-of-thermodynamical-work or 
proofs-of-position, in an unconditional way from the second law of thermodynamics in the 
version of Landauer's principle, stating that the erasure of information has a thermodynamic
price in the form of free energy to be turned into heat dissipated into the environment.
Specifically, our protocols are secure under the sole assumption that the free energy 
accessible to an opponent is limited. Interesting open questions are whether their information-theoretic security is universally composable (which can be delicate~\cite{vilasini2019composable,laneve2021impossibility}), and whether the bounds on the free-energy usage can be made tight by adding ``clean-up'' rounds to erase intermediary data (in analogy to~\cite{bennett2014quantum}).

Whereas our protocols represent theoretical proofs of principle, they are not practical for the 
time being, for several reasons, including the smallness of Boltzmann's constant, as well as
the current abundance of free energy in the Universe. The latter is not granted for all future. 
In this sense, and in a spirit related to the term ``post-quantum cryptography", we pursue a
(probably) longer-term goal one might like to call ``pre-heat-death cryptography".

\section*{Acknowledgment}
We thank Renato Renner and Charles Bennett for insightful discussions about the physicality of the model, and we thank Venkatesh Vilasini for interesting discussions about composability. We thank all three anonymous reviewers for their very helpful critiques and suggestions. This research was supported by the Swiss National Science Foundation (SNF). 

\bibliographystyle{ieeetr}
\bibliography{../../ddd_all}

\end{document}